\documentclass[aps,prl,reprint,groupedaddress,superscriptaddress,showpacs,floatfix]{revtex4-1}
\usepackage{braket}
\usepackage{graphicx}
\usepackage{dcolumn}
\usepackage{bm}
\usepackage{epsfig}
\usepackage{amsmath}
\usepackage[usenames,dvipsnames]{color}
\usepackage{booktabs}
\usepackage{color}
\usepackage{hyperref}
\begin{document}

\title{Fermi surface topology and non-trivial Berry phase in the flat-band semimetal Pd$_3$Pb.}

\author{Mojammel A. Khan}

\email{mkhan19@anl.gov}
\affiliation{Materials Science Division, Argonne National Laboratory, 9700 South Cass Avenue, Argonne, Illinois 60439, USA}

\author{Po-Hao Chang}
\affiliation{Department of Computational and Data Sciences, George Mason University, 4400 University Dr. Fairfax, Virginia 22030, USA}

\author{Nirmal Ghimire}
\email{Current address: Department of Physics and Astronomy, George Mason University, 4400 University Drive,
Fairfax, Virgina 22030, USA}
\affiliation{Materials Science Division, Argonne National Laboratory, 9700 South Cass Avenue, Argonne, Illinois 60439, USA}

\author{Terence M. Bretz-Sullivan}
\affiliation{Materials Science Division, Argonne National Laboratory, 9700 South Cass Avenue, Argonne, Illinois 60439, USA}

\author{Anand Bhattacharya}
\affiliation{Materials Science Division, Argonne National Laboratory, 9700 South Cass Avenue, Argonne, Illinois 60439, USA}

\author{J. S. Jiang}
\affiliation{Materials Science Division, Argonne National Laboratory, 9700 South Cass Avenue, Argonne, Illinois 60439, USA}

\author{John Singleton}
\affiliation{National High Magnetic Field Laboratory, Los Alamos National Laboratory, MS-E536, Los Alamos, New Mexico 87545, USA}

\author{J. F. Mitchell}
\affiliation{Materials Science Division, Argonne National Laboratory, 9700 South Cass Avenue, Argonne, Illinois 60439, USA}

\date{\today}
\begin{abstract}
A study of the Fermi surface of the putative topological semimetal Pd$_3$Pb has been carried out using Shubnikov-de Haas (SdH) oscillations measured in fields of up to 60 T.  Pd$_3$Pb has garnered attention in the community due to a peculiar Fermi surface that has been proposed theoretically by Ahn, Pickett, and Lee, [Phys. Rev. B 98, 035130 (2018)] to host a dispersion-less band along $X-\Gamma$ as well as multiple triply-degenerate band crossings that, under the influence of spin-orbit coupling, lead to ten four-fold degenerate Dirac points. Analysis of the SdH oscillation data verifies the calculated multi-sheet Fermi surface, revealing a $\Gamma$ centered spheroid that had not been resolved experimentally in prior studies. A comprehensive, angle-dependent analysis of the phase of the SdH oscillations convincingly demonstrates a non-trivial Berry phase for two bands along $\Gamma-R$, supporting the theoretical predictions, while simultaneously evidencing interference between extremal orbits that mimics a trivial Berry phase at intermediate angles.
\end{abstract}

\maketitle
\section{Introduction}
Crystalline solids have emerged as novel testbeds in the search for exotic fundamental particles~\citep{yan2012topological,ando2015topological}; for example, three-dimensional topological semimetals offer a chance to explore new topological phases beyond those predicted in high-energy physics~\citep{PhysRevX.6.031003,bradlyn2016beyond,wang2016hourglass}. Along this journey, many new phases have been discovered, including skyrmions~\citep{muhlbauer2009skyrmion}, Weyl fermions~\citep{topo.review.weyl}, and Majorana fermions~\citep{topo.review.TSC}. These discoveries not only access the fundamental physics of unusual particles but also strongly suggest the application of such exotic states to future technologies such as spin-based electronics and quantum computation.

Recently, the cubic compound Pd$_3$Pb, has been proposed~\citep{pd.theory} to host a novel triple point (TP) fermionic phase with threefold degeneracies along certain high symmetry lines. In addition to the triple points, Pd$_3$Pb also exhibits a dispersionless band along the $\Gamma-X$ line that lies close to the Fermi energy ($E_{\rm F}$). Band-structure calculations in the absence of spin-orbit coupling (SOC) show a combination of triple nodal points and three-dimensional nodal rings, giving rise to topological surface states along the $\Gamma-R$ and $R-M$ lines. The inclusion of SOC produces ten fourfold-degenerate Dirac points as well as topological character on the $k_{\rm z}$=0 plane. 
  
Recent transport studies of Pd$_3$Pb have revealed a large non-saturating magnetoresistance and a high mobility, $\mu\approx$2$\times$10$^3$ cm$^2$ V$^{-1}$ s$^{-1}$.~\citep{Pd.first}. In another study, torque magnetometry has been utilized  to explore the Fermi surface (FS), which shows the existence of several electron and hole pockets with small carrier masses~\citep{Pd2}. However, a complete picture of the FS topology including the topological aspects of the bands is yet to be reported. Moreover, the crystals used in torque studies were found to be doped to an equivalent of 3\% Bi-substitution on the Pb site, placing $E_{\rm F}\approx$50~meV above that calculated by Density Functional Theory (DFT)~\citep{Pd2}. It was asserted that $\approx$1\% Bi-doping is required to place $E_F$ precisely on the flat band~\citep{pd.theory}. In this study, we use quantum oscillations measured on oriented single crystals of Pd$_3$Pb to complete an experimental mapping of the FS, revealing a previously unresolved hole pocket at $\Gamma$. From the frequency dependence on the position of $E_F$, we find our crystals to be slightly electron doped, placing the $E_F\approx$30 meV above calculated value of $E_F$ for nominal Pd$_3$Pb. Importantly, we show that the bands between $\Gamma-R$ and $R-M$ give rise to non-trivial Berry phase, validating the topological features identified by theory~\citep{pd.theory}. 

Single crystals of Pd$_3$Pb were grown using a self flux method described elsewhere~\citep{Pd.first}. Shubnikov-de Haas (SdH) oscillations were measured in both low field (up to 14 T in a Quantum Design Physical Property Measurement System (PPMS)) and at high field in a 65 T magnet at the National High Magnetic Field Laboratory (NHMFL) Pulsed-Field Facility ~\citep{singleton2004national}. A detailed description of experimental methods are provided in the Supplemental Materials (SM)~\citep{supp}~\nocite{PDO.1,PDO.2,PDO.3,ho2007haas,vasp,gga}.

\section{Results and Discussions}

\begin{figure*}[]
\centering
\includegraphics[width=1.0\textwidth]{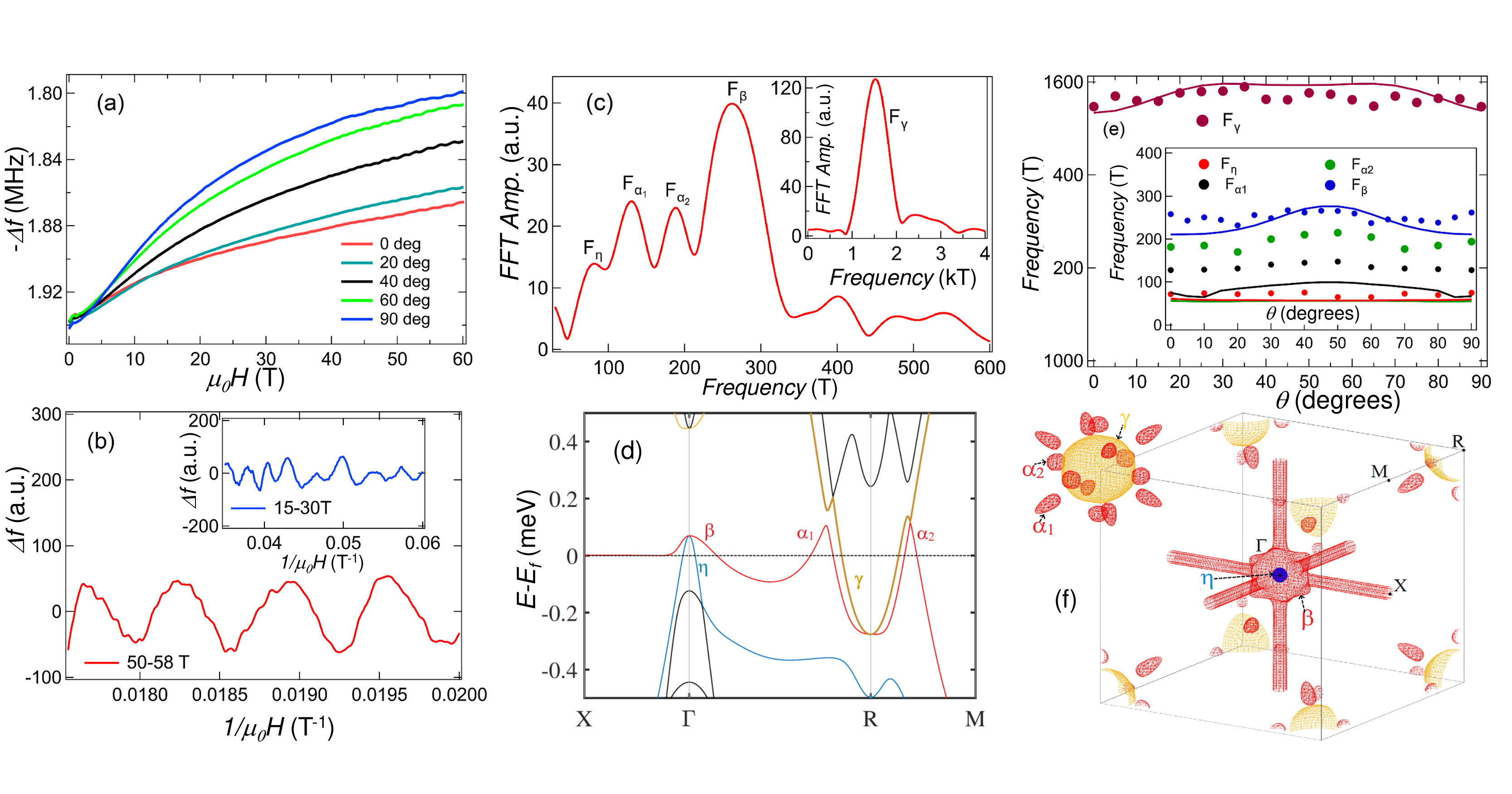}
\caption{Shubnikov-de Haas oscillation and Fermi surface of Pd$_3$Pb. (a) Raw oscillations shown as negative of change in resonant frequency of PDO coil. (b) PDO data shown in two field ranges after a smooth background subtraction : Low field, 15-30T (Inset) and high field range, 50-58 T. (c) fast Fourier transform (FFT) of the oscillation data in two different field ranges. The low and high frequencies are evident. Inset: FFT showing the 1570 T frequency. (d) Band structure of Pd$_3$Pb with spin-orbit coupling expanded over a small energy range to show the electron and hole pockets that are present in the FFTs. (e) Angular dependence of the frequencies identified by FFTs. The solid lines are the theoretical values as described in the text. (f) Fermi surface of nominal Pd$_3$Pb projected into the first Brillouin zone. The color of the bands are same as for panel (d). Description of the band structure calculations can be found in SM~\citep{supp}.}\label{figure 1}
\end{figure*}

Fig.~\ref{figure 1}a shows the SdH oscillations measured through a change in resonant frequency, $\Delta f$, and hence the magnetoresistance (MR), since $\Delta$\textit{f}$\propto$-$\Delta\rho$~\citep{PDO.1}. The increasing MR is consistent with previous reports~\citep{Pd.first}. Upon subtraction of a smooth background, the oscillations become more pronounced, as shown in Fig.~\ref{figure 1}b, where the field is applied along the $a$-axis. Here the data were divided into two field regimes. The high frequency oscillations are apparent above 50 T. In the field range from 10-30 T, a fast Fourier transform (FFT) reveals several low frequencies with broad maxima. This is a consequence of weaker low frequency oscillations in the high field data. The maximum of the FFT peak was also verified via PPMS measurements (see SM~\cite{supp} Fig. S1). The results of the FFT are shown in Fig.~\ref{figure 1}c with frequencies, $F_{\alpha_1}$= 128 T, and $F_{\alpha_2}$ = 182 T, and $F_\beta$ = 258 T, while the high frequency oscillations yield $F_\gamma$ = 1570 T. The observed frequencies are in good agreement with previous reports~\citep{Pd.first,Pd2}. However, we are also able to discern an additional frequency $\eta$ with $F_\eta$ = 72 T, that was not reported in Ref.~\citep{Pd2}. The absence of the additional frequency in the torque data can be understood either from the spherical shape of the corresponding FS sheet or the dependence of $F_\eta$ on the position of $E_{\rm F}$ (see SM~\citep{supp} for more discussion).

The observed frequencies can be related to the calculated band structure shown in Fig.~\ref{figure 1}d for nominal Pd$_3$Pb. Three hole pockets, $\eta$, $\beta$, and $\alpha_1$ are present along $\Gamma$-$R$; a large electron pocket $\gamma$ is centered at $R$, followed by another hole-pocket, $\alpha_2$, along $R$-$M$. The nontrivial nature of the carriers in this band arises from Dirac crossings near $E_F$ along $R-M$ and $R-\Gamma$ lines. These crossings evolve from the triple nodal points under the influence of spin-orbit coupling, as noted in Ref.~\citep{pd.theory} and can be seen in Fig.~\ref{figure 1}d. Fig.~\ref{figure 1}e shows the angular dependence of the observed frequencies with solid lines indicating calculated values with $E_F$ at 30 meV (see SM~\citep{supp} Fig.S3). All frequencies are well accounted for within a full 90 degree rotation, establishing the three dimensional nature of the FS sheets. This is also evident from the FS plots in Fig.~\ref{figure 1}f, where the individual FS sheets are projected into the first Brillouin zone for undoped Pd$_3$Pb. These results are consistent with previous report~\citep{pd.theory}. The $\eta$ and $\gamma$ FS sheets are nearly spherical, while $\alpha_1$ is highly anisotropic with a carrot-like shape and $\alpha_2$ is anisotropic and elliptical. The $\beta$ sheet takes a cube-like shape and encloses $\eta$. The observed frequencies and the calculated values are listed in the Table S1 of~\citep{supp}. Overall, theoretical estimates are in reasonably good agreement with experimental observations, except the frequencies $F_{\alpha_{1,2}}$ and $F_\eta$ are underestimated, similar to the earlier report~\citep{Pd2}. 

\begin{figure}[]
\centering
\includegraphics[width=0.45\textwidth]{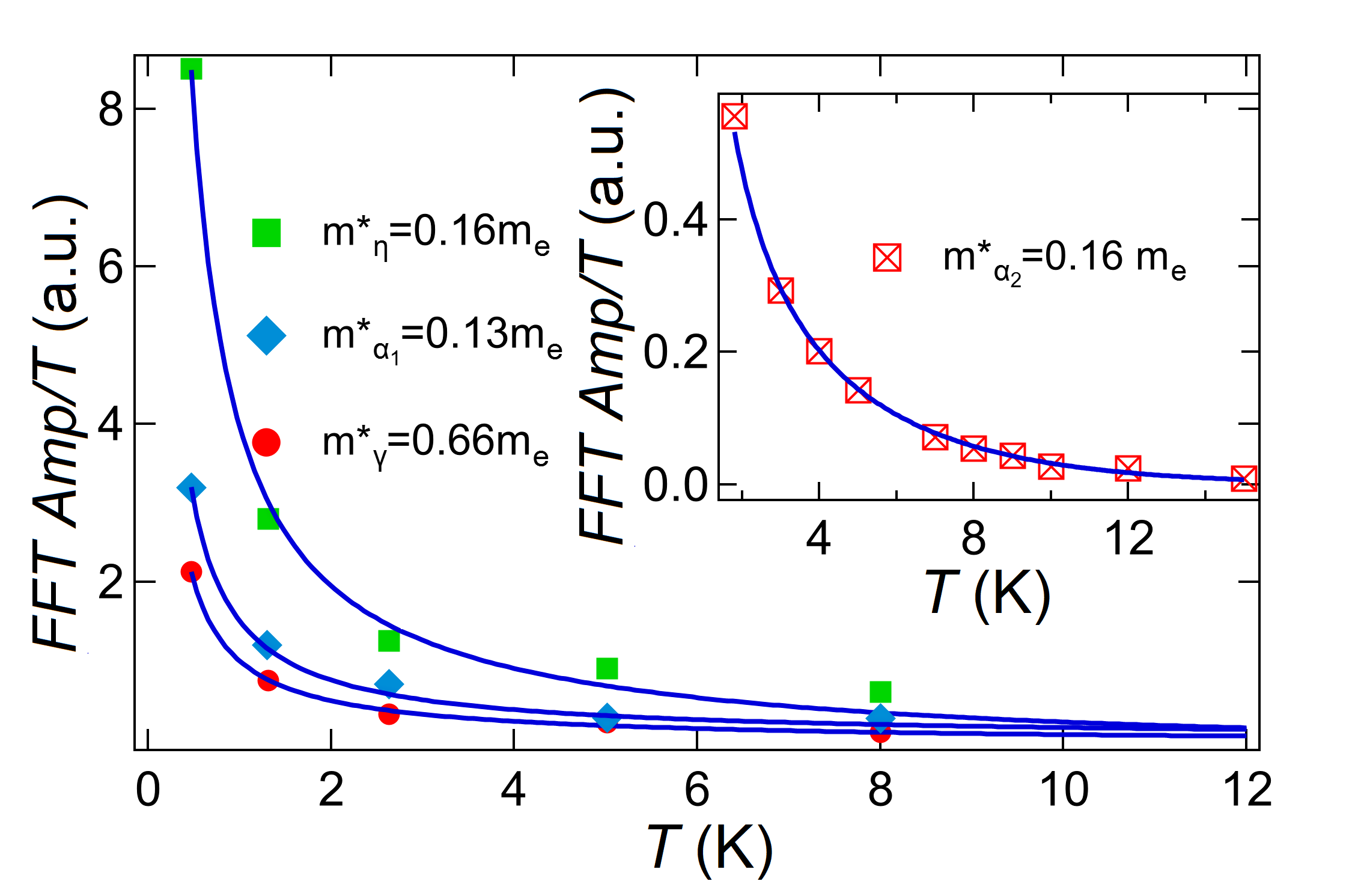}
\caption{Band effective mass. LK fit to the temperature dependent amplitude. Inset: LK fit to the 185 T frequency data. Temperature dependence of the FFT peak amplitudes are given in Fig. S1 of the SM~\citep{supp}.}\label{mass}
\end{figure}

From the temperature dependence of the amplitude of the oscillations, we have determined the effective mass of the carriers of the bands $\alpha_{1,2}$, $\eta$, and $\gamma$ (Table S1). For the smaller frequencies, we utilized a combination of data up to 14 T from the PPMS and up to 30 T from NHMFL, and for the larger frequency, $F_\gamma$, we used the data taken at NHMFL. As shown in Fig.~\ref{mass}, the fit yields $m*_{\alpha1}\approx0.13(2)m_e$, $m*_{\alpha2}\approx0.18(4)m_e$, $m*_{\eta}\approx0.16(2)m_e$, and $m*_{\gamma}\approx0.66(3)m_e$. The experimental values are in excellent agreement with the theoretical estimates. For the $\beta$ band, the FFT peak amplitude decreases rapidly (see Fig. S1 of~\citep{supp}), preventing similar LK fits from extracting the effective mass reliably. This indicates a heavier effective mass for this band, consistent with the calculated value of $\approx 1.3m_e$.

The $\pi$-Berry phase accumulated in the cyclotron
motion of the quasiparticles is a fundamental property of topological semimetals~\citep{burkov2016topological}. The effect of this additional phase factor on quantum oscillations can be studied via the LK formalism developed for a three dimensional (3D) system with arbitrary band dispersion~\citep{Berry.phase.TaP,Berry.phase.2,Berry.phase.3,BiPd.me};

\begin{equation}
\Delta\rho\propto\left( \frac{B}{2F}\right)^{1/2}R_T R_{\rm D} R_{\rm S} \cos\left[2\pi\left\lbrace \frac{F}{B}+\gamma-\delta\right\rbrace \right] 
\end{equation}\label{LK eq}
Here, $R_{\rm S} = \cos(\frac{p\pi g m^*}{2})$, the spin damping factor, \textit{p} is the harmonic index, \textit{g} is the \textit{g}-factor. $F$ is the SdH frequency, and the Berry phase, $\phi_{\rm B}$, is related via the phase factor, $\mid\gamma$-$\delta\mid$, where $\gamma = \frac{1}{2}$-$\frac{\phi_B}{2\pi}$, and $\delta$ is related to the dimension of the Fermi pocket with values 0 for two dimensional (2D) and $\pm$1/8 for 3D cases(+ for minimal and - for a maximal cross section of the constant energy surface)~\citep{BiPd.me}. Thus, values for $\mid\gamma$-$\delta\mid$, with a nontrivial $\pi$-Berry phase are 0 for 2D and 1/8 for 3D Fermi surfaces. If the value of \textit{gm*}$>>$2, $R_{\rm S}$ can impart a phase shift in the oscillations, although the Berry phase may be zero. In many d-electron systems, the value of $g$ is large~\citep{Berry.phase.3}, for example, $g \approx$3 in AuBe~\citep{AuBe}. In Pd$_3$Pb, the absence of a prominent second-harmonic in our SdH data (cf.~\citep{AuBe}; see Fig.~\ref{figure 1}c) coupled with the small $m^{*}$ values of the SdH frequencies under consideration suggest that $R_{S}$ can be neglected safely and that any phase change in our oscillation data is attributable to the effect of a non-trivial Berry phase~\citep{Berry.phase.3,AuBe}.

\begin{figure*}[]
\centering
\includegraphics[width=0.8\textwidth]{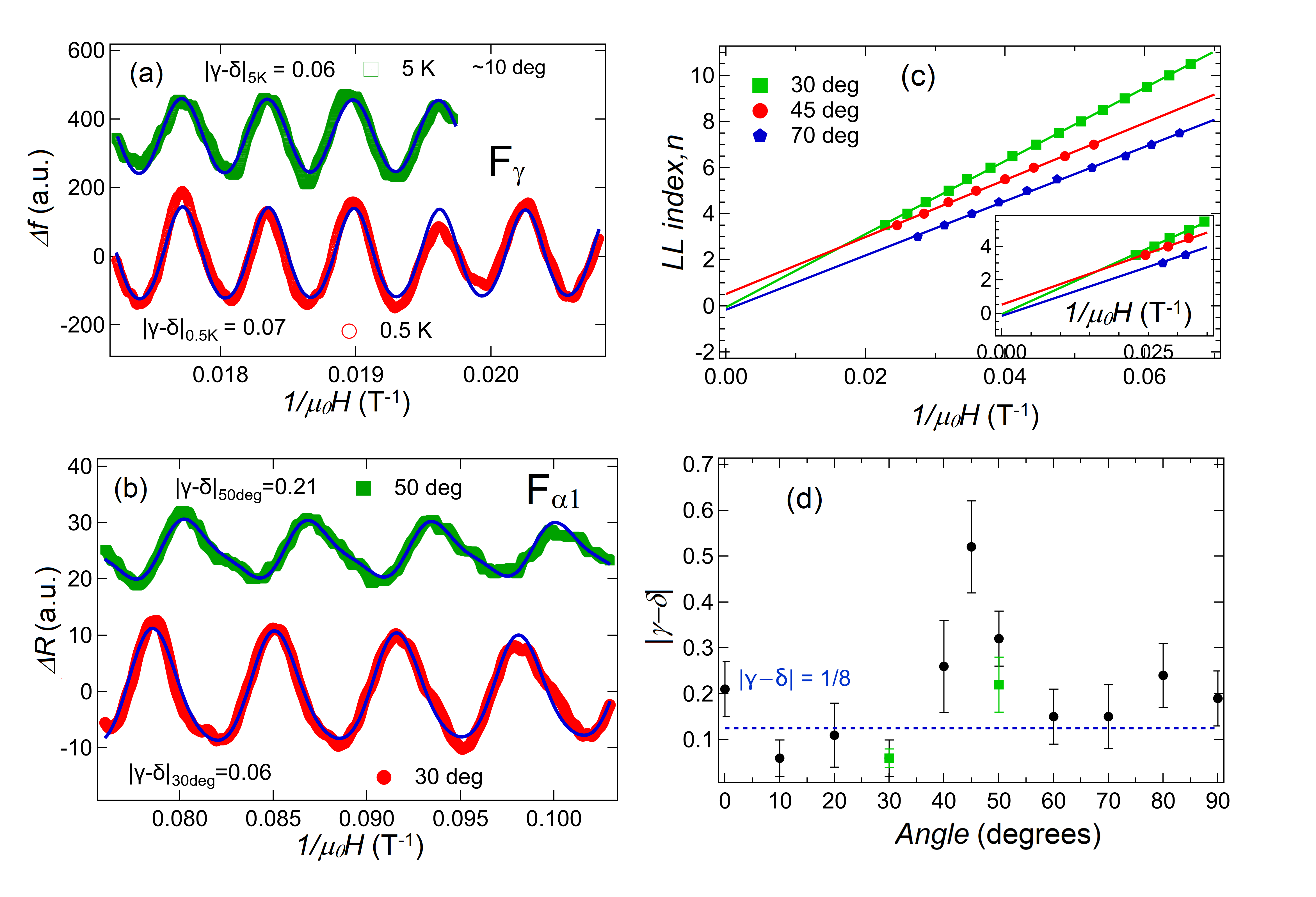}
\caption{Berry phase analyses for $F_{\alpha_1}$ and $F_\gamma$. (a) A single band LK fit (blue solid line) to the oscillation data for $F_\gamma$ at 0.5 K (red open circle) and 5 K (olive squares) measured at NHMFL. (b) Single band LK fit (blue solid line) to the oscillation data for $F_{\alpha_1}$ at 30 degrees (red solid circles) and 50 degrees (olive solid squares) and 2 K measured in the PPMS. One higher order harmonic (2$F$) and the fundamental frequency were used for the LK fits. (c) LL-fan diagram for $F_{\alpha_1}$ as described in the text. Inset: Expanded view near the origin. (d) Angular dependence of the phase factor for $F_{\alpha_1}$ estimated from the LL-fan plots. At 30 and 50 degrees, values derived from LK-fits are plotted and are shown as green solid squares. }\label{phase}
\end{figure*}

Using the LK formula, we studied two dominant frequencies, $F_{\alpha_1}$ and $F_\gamma$, which arise from the FS sheets that lie along $\Gamma-R$ and $R-M$ and are predicted to be topological~\citep{pd.theory}. Due to the complex FS, we separated the data ranges where the individual frequencies are dominant. For $F_\gamma$, it is evident that this frequency is dominant above 45 T, while for $F_{\alpha_1}$ the relevant field range is below 25 T. The low field data are complicated by other frequencies that contribute to oscillations in varying strengths. From the FFT in low field range($\approx$9-13 T) at 2 K, we find that when the field was along 30 and 50 degrees from the crystallographic $a$-axis, $F_{\alpha_1}$ dominates and can be used for LK fit. Fig.~\ref{phase}a shows a single band LK fit to the oscillation data at 0.5 K and 5 K for the high frequency, $F_\gamma$, in the field range 48--58 T with field applied $\approx$10 degrees from the $a$-axis. At 0.5 K, the fit yields $\mid\gamma$-$\delta\mid$=0.07$\pm$0.03 indicating a non-trivial topology with a $\pi$-Berry phase. From the fit, $F$=1572$\pm$4 T, consistent with the value we find via FFT, the effective mass, $m^*$=0.65 m$_e$, similar to the fits in Fig.~\ref{mass}, and the Dingle temperature, $T_{\rm D}$= 35$\pm$3K similar to that reported (42 K) in Ref.~\citep{Pd2}. 
At high fields and low temperatures, the Zeeman splitting between the spin-up and spin-down Fermi surfaces may be noticeable as a pronounced second harmonic in the SdH oscillations, potentially leading to difficulties in attributing a reliable phase. As discussed above, since we do not observe such harmonics, the spin-splitting term can be neglected in the LK fit. However, for completeness, we have used the data at 5 K and performed the LK fit as seen in Fig.~\ref{phase}a. At higher temperature any residual effect of Zeeman splitting will be reduced by the thermal broadening of the Fermi-Dirac distribution function. The fit yields $\mid\gamma$-$\delta\mid$ $\approx$0.06$\pm$0.02, in agreement with that found at 0.5 K and confirming the non-trivial Berry phase for this band.

A single band LK fit to the low field data at 2 K for $F_{\alpha_1}$ is shown in Fig.~\ref{phase}b. 
At 30 degrees, $\mid\gamma$-$\delta\mid$=0.06$\pm$0.02, indicating a $\pi$-Berry phase. Moreover, $F$, and $m^*$ agree with the values extracted from the FFT and thermal damping fits discussed earlier. The Dingle temperature, $T_{\rm D}\approx$11$\pm$3 K is somewhat smaller than reported in Ref.~\citep{Pd2}(25 K). 
Interestingly, the LK fit to the data at 50 degrees (Fig.~\ref{phase}b), yields larger $\mid\gamma$-$\delta\mid$=0.22$\pm$0.06, indicating an apparent change in the phase with angle. The deviation of the phase factor is also signaled through the shape of the oscillations at this angle, which become saw-tooth like rather than purely sinusoidal. 

To explore this behavior, we have investigated the angular dependence of the Berry phase for $F_{\alpha_1}$ between 0-90 degrees by rotating the crystal about [0 1 0] with 0 deg $\approx$ [1 0 0] and 90 deg $\sim$ [0 0 1] and by plotting the oscillation maxima ($n$) and minima ($n$+1/2) versus corresponding inverse field values (LL fan diagrams)~\citep{LL.fan}. For this, we have used the data taken at NHMFL up to 52 T, with LL level $n$=3, and extracted the $F_{\alpha_1}$ oscillations from the data via a band-pass filtering process~\citep{BiPd.me}. The oscillation minima and maxima were then indexed and plotted as seen in Fig.~\ref{phase}c for three different angles (fits for rest of the angles are shown in Fig. S4 of~\citep{supp}). Here, a linear extrapolation of the data provides the phase factor from the intercept, plotted in Fig.~\ref{phase}d, while the slope is equal to the frequency of the oscillations and provides a consistency check on the filtering process as well as the estimation of the phase. The horizontal lines in Fig.~\ref{phase}d mark the value where the Berry phase for the 3D case is non-trivial (1/8). It is evident that $\mid\gamma$-$\delta\mid\approx1/8$, except for a sharply peaked region centered at 45 degrees.

Similar angular dependent changes in Berry phase  have been reported and attributed to a change in the topological state of the band to a trivial phase. Explanations for this topological phase transition include a quantum phase transition~\citep{Berry.phase.trivial,PhysRevLett.115.226401}, a spin zero effect that leads to a new topological phase~\citep{wang2018vanishing}, or an interaction-induced spontaneous mass generation of Dirac fermions~\citep{Berry.phase.trivial2}. However, in the present case, the maximum value at 45 degrees does not cross the value 5/8 to indicate a phase transition, and a negligible $R_{\rm S}$ term excludes the possibility of spin-zero effect. 
The change in phase factor could also arise from the change in the curvature of the Fermi surface, reflected through $\delta$~\citep{Berry.phase.2,PhysRevLett.113.246402,Berry.phase.trivial,Berry.phase.trivial2,Berry.phase.trivial3,PhysRevLett.113.246402}. However, within the precision of the DFT calculations, a change in curvature with respect to the position of the magnetic field for $F_{\alpha_1}$ can not be resolved. 

Based on the non-sinusoidal appearance of the oscillations, we suggest that the apparent phase variation is an artifact due to changes in the relative amplitudes of several series of SdH oscillations that are closely spaced in frequency. This effect may result from the 3D shape of the $\alpha_1$ FS sheet and its position close to the $\gamma$ sheet. As the crystal is rotated from the high symmetry direction (i.e., $a$-axis) close to 45 degrees, the different $\alpha_1$ sheets (see Fig.~\ref{figure 1}f) will possess slightly different extremal orbits, resulting in a smearing of the observed oscillations due to the presence of several similar frequencies, and rendering the phase difficult to extract. Moreover, close to this angle, magnetic breakdown could occur due to the tunneling of carriers between $\alpha_1$ and $\gamma$~\citep{Berry.phase.3}. As the field grows, this will reduce the fraction of quasiparticles completing orbits about these $\alpha_1$ sheets, attenuating the corresponding SdH oscillations, while leaving those not subject to breakdown unaffected. Support for this explanation is provided by the unexpectedly large Dingle temperatures found for the oscillations associated with $\gamma$ ($\approx 35$~K) and $\alpha_1$ ($\approx 11$~K) in this angular range. However, the FFT data at 45 degrees (see SM~\citep{supp} Fig. S5) do not reveal clear evidence of sum or difference frequencies due to the proposed breakdown orbits. It is possible that such oscillations may be relatively weak due to a relatively large effective mass associated with the breakdown orbit. 

\section{Conclusion}
In summary, we have studied the Fermi surface topology of the semimetal Pd$_3$Pb with a combination of low- and high-field measurements and validated the topological nature of this compound predicted by theory. We have experimentally mapped the complete FS, including a hole pocket that had previously been unresolved. Corroborating recent calculations~\citep{pd.theory}, we find evidence of $\pi$-Berry phase and hence non-trivial topology for two bands that lie along $\Gamma-R$ and $R-M$ lines. Finally, we point out that the study of the angular variation of the phase factor is important for the systems with complex Fermi surfaces. Incorrect assignment of trivial or non-trivial Berry phases can result from the analysis of the oscillations with field along a single directions.

\section{acknowledgments} 
This research was sponsored by the U.S. Department of Energy, Office of Science, Basic Energy Sciences, Materials Sciences and Engineering Division. P.C. acknowledges support from the U.S. Department of Energy grant DE- SC0014337. A portion of this work was performed at the National High Magnetic Field Laboratory, which is supported by National Science Foundation (NSF) Cooperative Agreement Nos. DMR-1157490 and 1644779, the State of Florida and the U.S. Department of Energy (DOE) through the Basic Energy Science Field Work Proposal ``Science in 100 T.'' We thank Ivar Martin and Ulrich Welp for useful discussions.

\pagebreak
\widetext
\begin{center}
\textbf{Supplemental Materials for: Fermi surface topology and non-trivial Berry phase in the flat-band semimetal Pd$_3$Pb.}\\

\emph{Mojammel A. Khan, Po-Hao Chang, Nirmal Ghimire, Terence M. Bretz-Sullivan, Anand Bhattacharya, J.S. Jiang, John
Singleton, and J.F. Mitchell}
\end{center}
\setcounter{equation}{0}
\setcounter{figure}{0}
\setcounter{table}{0}
\setcounter{page}{1}
\makeatletter
\renewcommand{\theequation}{S\arabic{equation}}
\renewcommand{\thefigure}{S\arabic{figure}}
\renewcommand{\thetable}{S\arabic{table}}
\renewcommand{\bibnumfmt}[1]{[S#1]}
\renewcommand{\citenumfont}[1]{S#1}


\section{}

\begin{figure*}[ht]
\renewcommand{\thefigure}{S\arabic{figure}}
\includegraphics[scale=0.5]{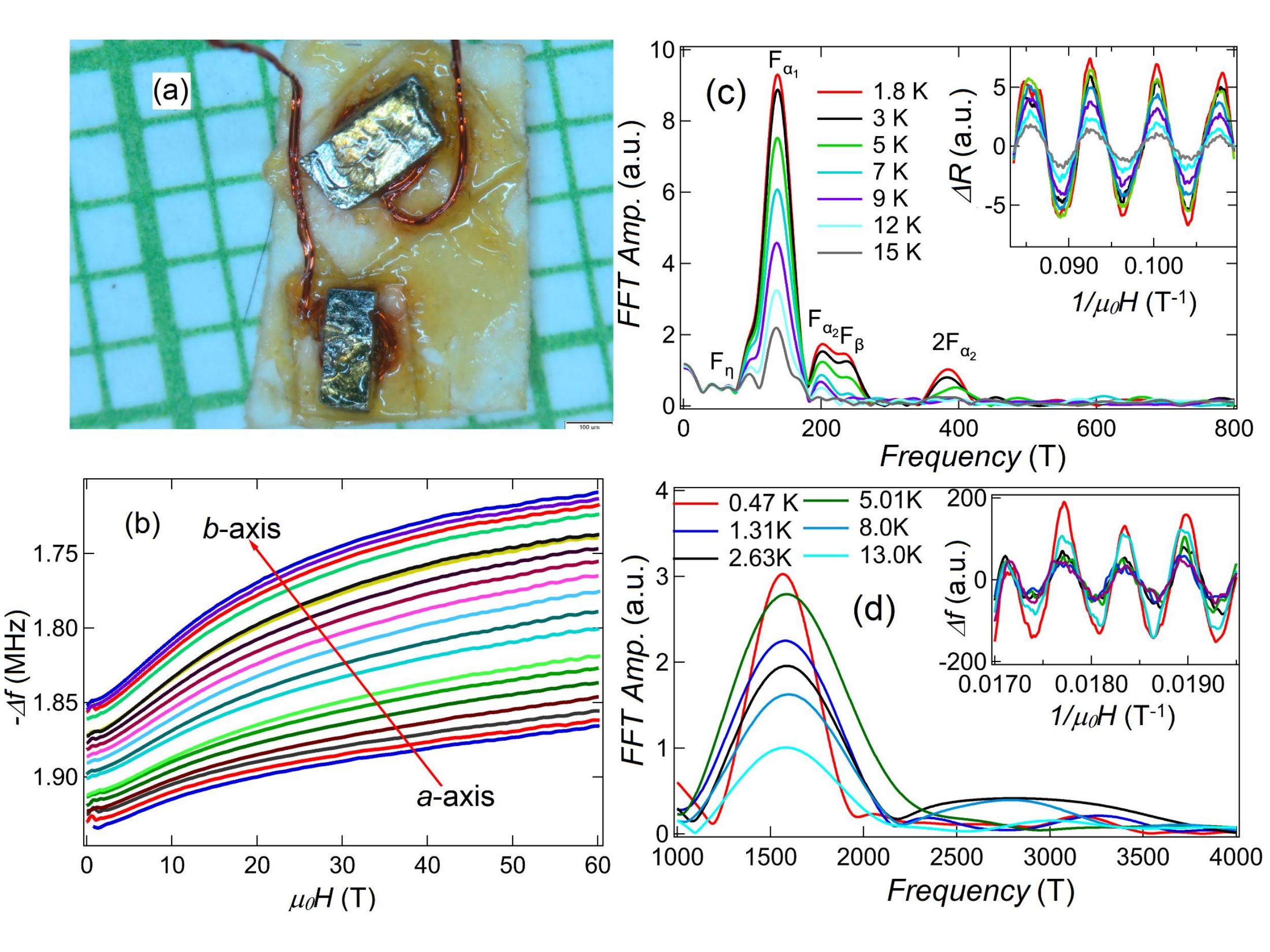}
\caption{(a) Two rectangular pieces of Pd$_3$Pb prepared for PDO measurements. (b) Raw SdH oscillations
measured as a change in resonant frequency, $\Delta f$. Crystal was rotated a full 90 degrees toward the \textit{b}-axis starting from field parallel to crystallographic a-axis. (c) Temperature dependence of the FFT amplitude in low field range taken in a PPMS. Inset: Background subtracted data showing the amplitude of oscillations decaying with increasing temperature. (d) Temperature dependence of the oscillation amplitude in the high field range (40-58 T). Inset: Background subtracted data used in FFT. }\label{figs1}
\end{figure*}

\begin{figure*}[ht]
\renewcommand{\thefigure}{S\arabic{figure}}
\includegraphics[scale=0.3]{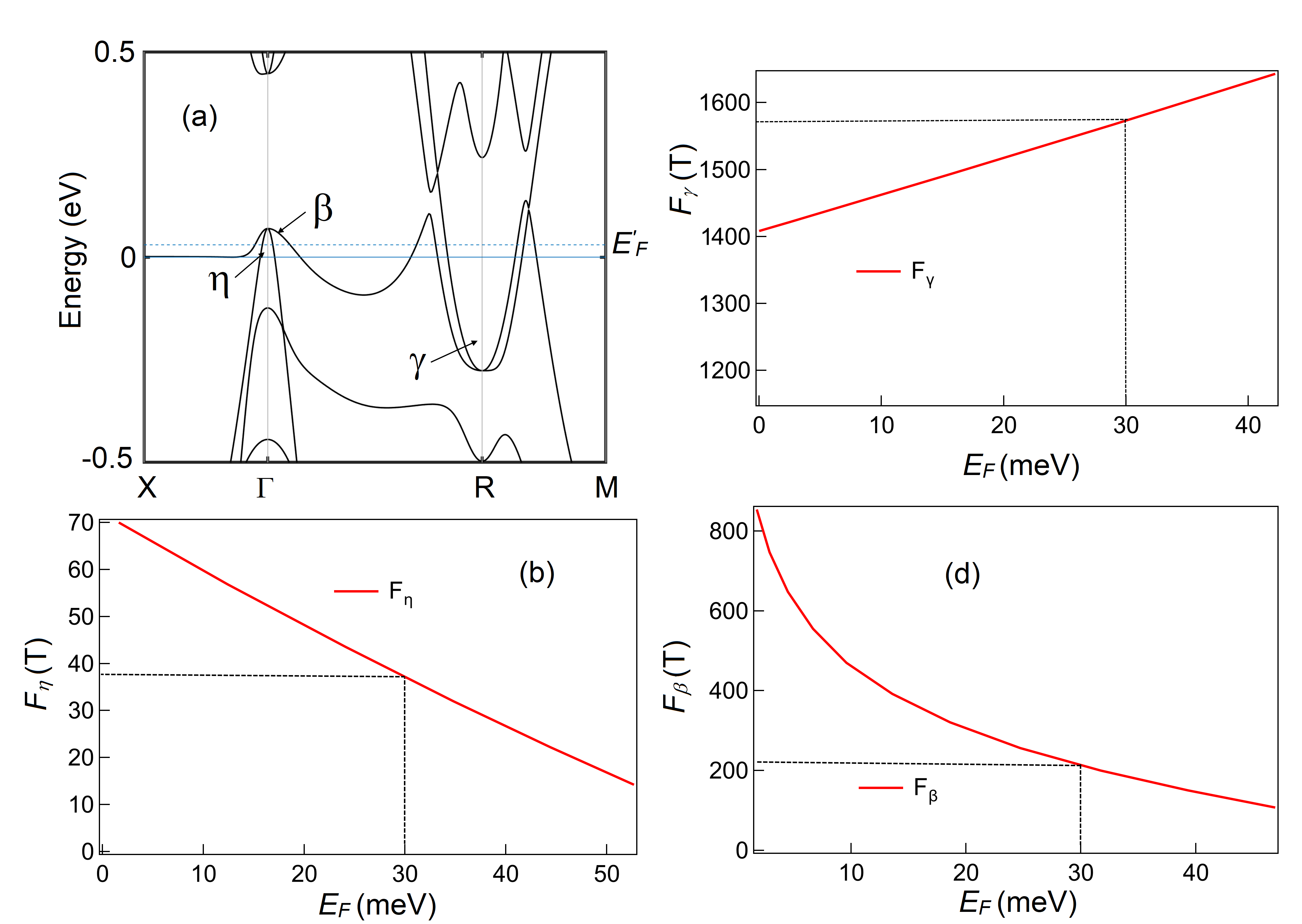}
\caption{(a) Band structure with $E_F$ at 0 meV (solid horizontal line) and at 30 meV (dashed horizontal line). (b-d) Frequency dependence on the position of $E_F$ for $F_\gamma$, $F_\eta$, and $F_\beta$, respectively. Dashed lines (b-d) mark $E_F$ calculated for $F_\gamma$, $F_\eta$, and $F_\beta$. }\label{figs2}
\end{figure*}

\begin{figure*}[ht]
\renewcommand{\thefigure}{S\arabic{figure}}
\includegraphics[scale=0.4]{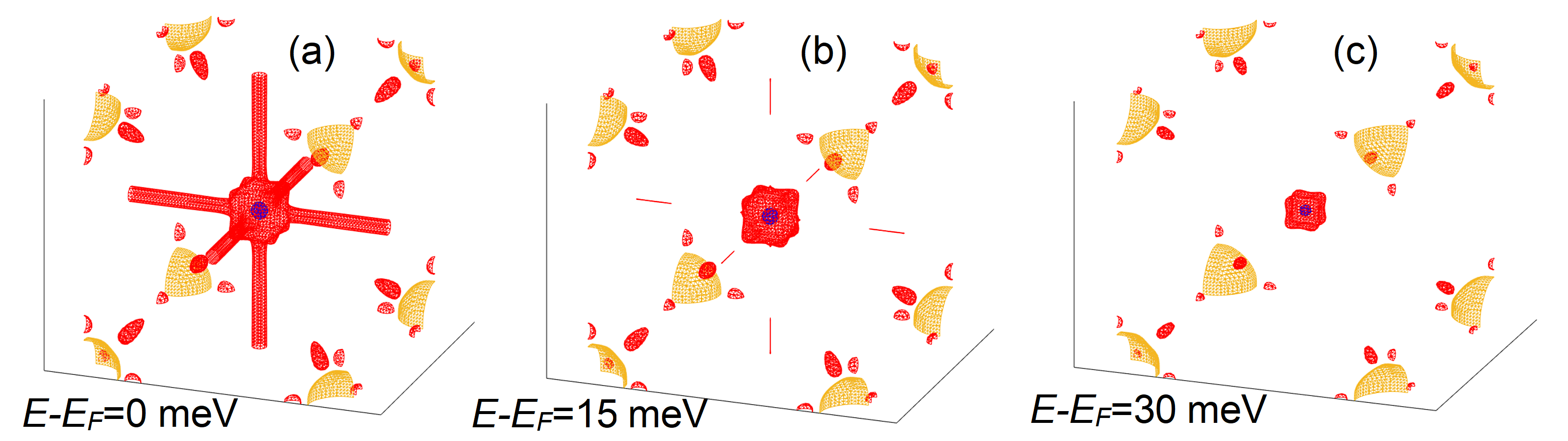}
\caption{Fermi surface with $E_F$ calculated at (a) 0 meV, (b) 15 meV, and (c) 30 meV. The long tubes which originate from the flat-band are absent at 30 meV, confirming the experimental determination of the FS.}\label{figs new FS}
\end{figure*}

\begin{figure*}[ht]
\renewcommand{\thefigure}{S\arabic{figure}}
\includegraphics[scale=0.2]{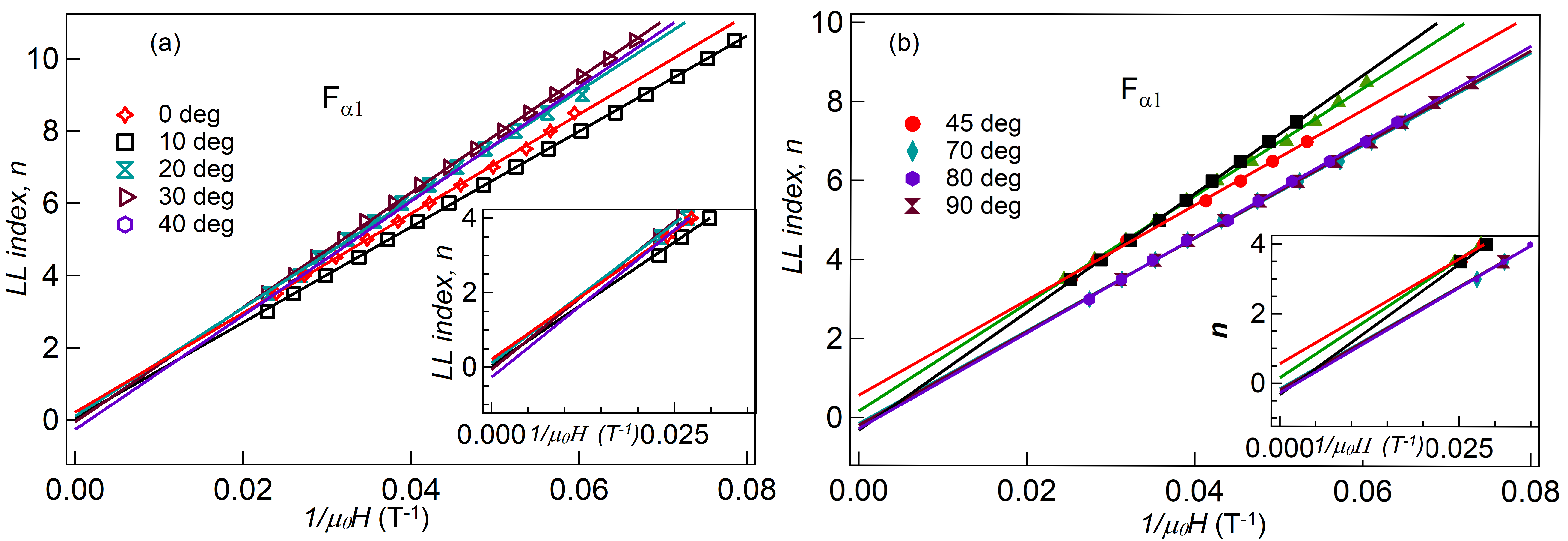}
\caption{(a) LL-fan diagram for $F_{\alpha 1}$ from 0-40 degrees and (b) 45-90 degrees. Solid lines are a linear fit to the data. Insets: View of the fits near the intercepts.}\label{figs3}
\end{figure*}

\begin{figure*}[ht]
\renewcommand{\thefigure}{S\arabic{figure}}
\includegraphics[scale=0.3]{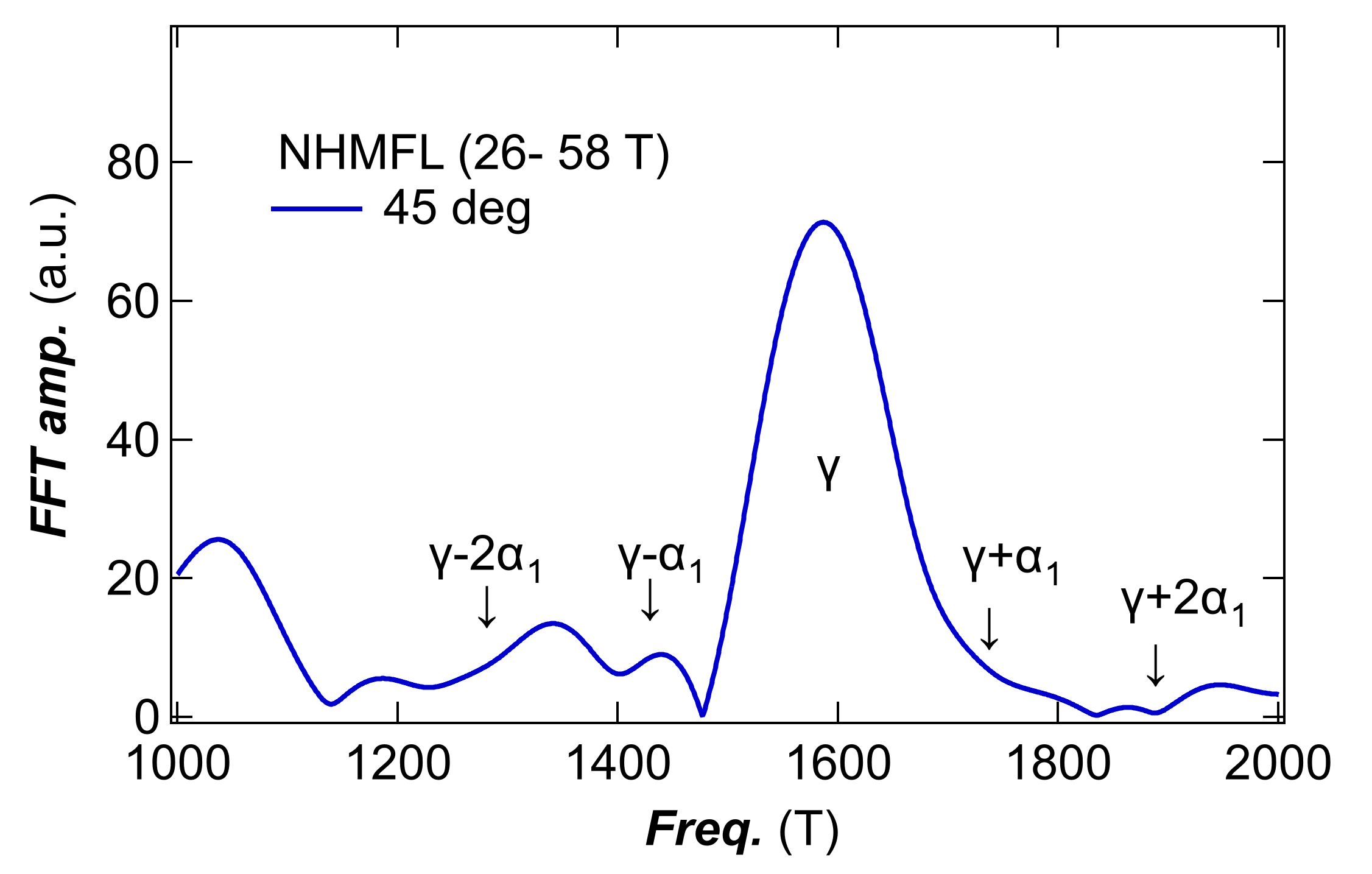}
\caption{ FFT of the high field data at 45 degree in a field range 26--58 T. The region around 1570 T frequency is shown for magnetic breakdown frequencies as described in main text. Arrows point to the positions of potential sum and difference frequencies.}\label{figs4}
\end{figure*}

A standard 4 probe resistance configuration was employed in the PPMS measurements. Electrical contacts were made by attaching 25 micron gold leads to a polished sample with H-20E epoxy. Measurement of Shubnikhov-de Haas oscillations was performed at the 65 T pulse-field facility at National High Magnetic Field Laboratory(NHMFL), Los Alamos National Laboratory. Here, a contactless conductivity method, utilizing a Proximity Detector Oscillator (PDO), was used to measure the oscillations via change in resonant frequencies of the oscillator circuit due to a change in the skin depth of the sample~\citep{PDO.1s}. A coil comprising 4--12 turns of 46-gauge high-conductivity copper wire was wound around the single crystal sample or, alternatively, the crystal was placed on top of the coil ( pancake style) as shown in Fig.~\ref{figs1}a. The number of turns employed depends on the cross-sectional area of the sample and is inversely proportional to the size of the sample~\citep{PDO.3s}. The coil forms part of a PDO circuit cavity that resonates at frequencies in the range of 22--29 MHz. By changing the temperature or by
applying a magnetic field, changes in the sample skin depth or differential susceptibility result~\citep{PDO.1s,PDO.2s}, changing the inductance of the coil, which in turn alters the resonant frequency of the circuit. The SdH oscillations are observed through this change in frequency. The signal from the PDO circuit is mixed down to $\approx$2 MHz to digitize the data prior to Fourier transformation.

Two samples in individual coils coupled to independent PDO apparatus were mounted
on a single-axis, worm-gear driven goniometer that controls the orientation of the samples with respect to applied field. The goniometer was then placed within a
$^3$He cryostat that provides temperatures down to 0.4 K. Temperature was measured using a Cernox sensor supplied and calibrated by Lakeshore Inc. Magnetic fields were deduced by integrating the voltage induced in an 11-turn coil, calibrated against the de Haas-van Alphen oscillations of the belly orbits of copper~\citep{ho2007haass}. The samples were rotated a full 90 degrees about [0 0 1] from crystallographic \textit{a}-axis to \textit{b}-axis.

Fig.~\ref{figs1}b shows the PDO data plotted as a negative change in the resonant frequency versus the applied field. Rising trends indicate the increase in magnetoresistance as described in the main text. In Fig.~\ref{figs1}c-d, we show the temperature dependence of oscillation data taken both in a PPMS and at NHMFL, respectively. Using the frequencies extracted from the low field data (Fig.~\ref{figs1}c) collected in the PPMS, the smaller frequencies extracted from the data collected at NHMFL was verified. Since the high field oscillations for smaller frequencies were weaker resulting in broader FFT peaks ( see Fig. 1c of the main text), PPMS data were used as a check to determine the exact frequency values. As seen in Fig.~\ref{figs1}c, smaller frequency $F_\eta$ can be observed which was not reported in earlier work done via torque magnetometry~\citep{Pd2s}. This can be understood from the fact that, torque magnetometry detects the anisotropic component of the magnetization, so that a spherical Fermi pocket may not produce an observable torque. Additionally, based on our DFT calculations, this FS sheet depends sensitively on $E_F$ (Fig.~\ref{figs2})  i.e., $F_\eta$ changes rapidly with electron concentration. In Ref.~\citep{Pd2s}, the authors estimated their samples to be electron doped equivalent to 3\% Bi, which is consistent with $E_{\rm F}$ raised by 54 meV relative to the calculated position in the undoped case. As discussed in the main text, all the FS sheet observed are 3D in nature. We note that no 1/cos$\theta$ dependence is observed in Fig. 1e of main text. Such behavior would be characteristic of 2D sheets such as the ``tubes'' radiating along $\Gamma-X$ directions in the calculated FS of undoped Pd$_3$Pb (Fig. 1f of main text). The self-doped nature of Pd$_3$Pb which results in higher position of $E_F$ explains the absence of this distinctive FS sheet in our measured SdH data(Fig. 1e of main text). In Fig.~\ref{figs new FS}, the FS at three different $E_F$ are shown, and it is evident that the long tubes are absent at 30 meV, supporting the experimental observation.

The amplitude of the SdH oscillations is described by the Lifshitz-Kosevich (LK) formalism with a thermal damping factor, $R_{T}$ = $\alpha T m^*/B \sinh(\alpha T m^*/B)$, while the field dependence of the amplitude is $R_{D}$ = exp(-$ \alpha T_{D} m^*/B)$. Here, $m^*$ is the effective mass of the quasiparticle in units of the electron mass, $\alpha$=2$\pi^2 k_{\rm B} m_e/e\hbar \approx 14.69$ T/K with $T$ in Kelvin and $B$ in Tesla, and $T_{D}$ is the Dingle temperature ($T_{D}$ = $\hbar/2\pi k_{\rm B} \tau_{\rm s}$). From the decay of the FFT peak amplitude, the effective masses of the band carriers were calculated and are presented in the main text, Fig. 2 and Table S1.

Density Functional Theory (DFT) calculations were performed using the projector-augmented wave (PAW) method, implemented in the Vienna Ab initio Simulation Package (VASP)~\citep{vasps} within the generalized gradient approximation (GGA) for the exchange-correlation potential~\citep{ggas}, and spin-orbit coupling (SOC) is included. A uniform 42$\times$42$\times$42 $k$-space mesh provided sufficient convergence for the Brillouin zone integration. 

\begin{table}
\renewcommand{\thetable}{S\arabic{table}}
  \caption{Fundamental frequencies, effective mass, $m^*$, and phase factor,  $\mid\gamma$-$\delta\mid$ for Pd$_3$Pb with field applied at 10 degrees from \textit{a}-axis. Some values are also compared with previous work~\citep{Pd2} as discussed in the text.}
  \label{tbl:excel-table}
  \includegraphics[width=0.6\textwidth]{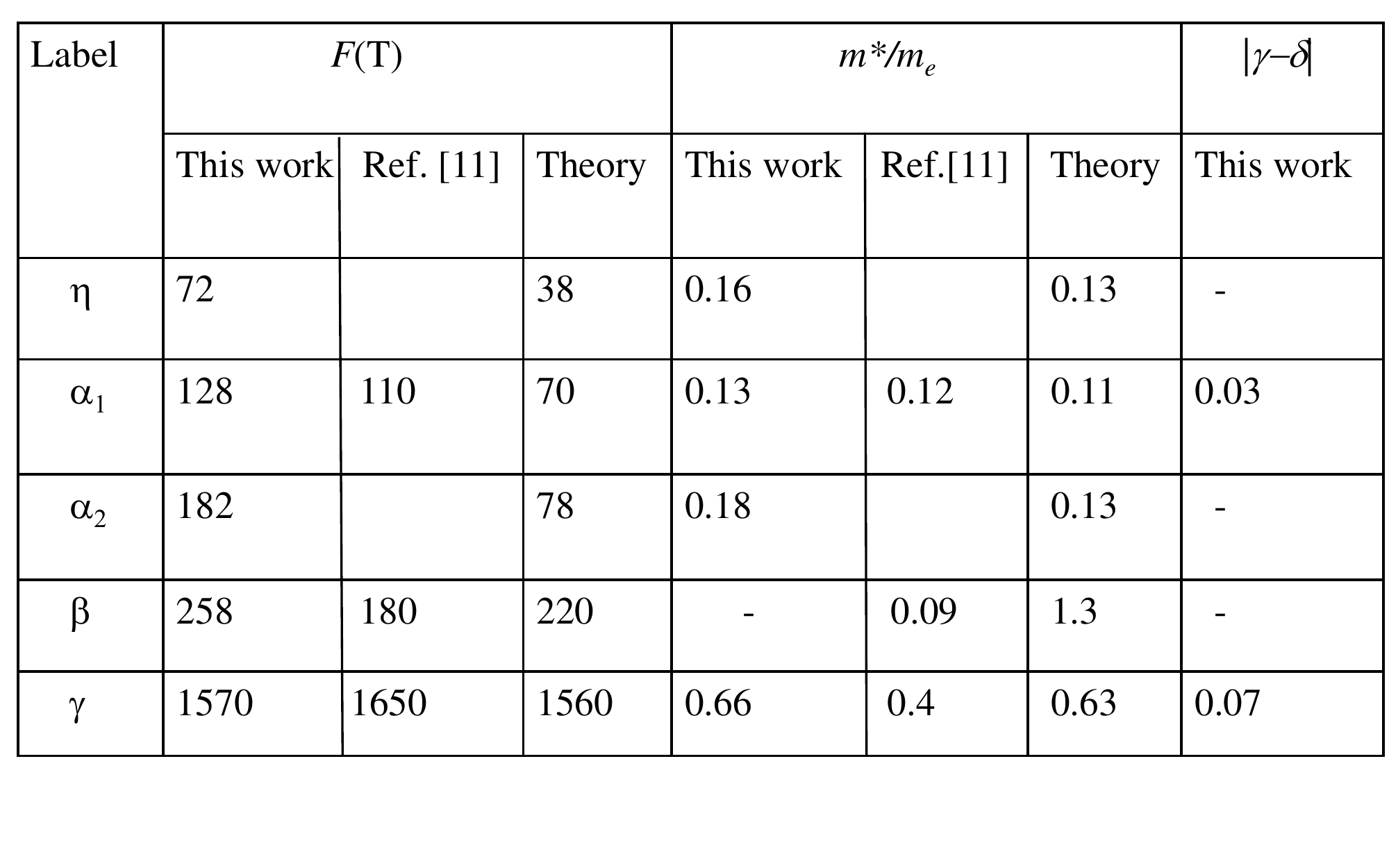}\label{fig-table}
\end{table}

In Fig.~\ref{figs2}, results of the band structure calculations are shown. When $E_F$ is raised to 30 meV, all the hole pockets are still present albeit with a smaller area, while the electron pocket at $R$-is the largest among all the FS pockets. The $E_F$ dependence of the two hole pockets at $\Gamma$ and the electron pocket at $R$ is shown in Fig.~\ref{figs2}(b-d). The $E_F$ dependent frequencies of three different closed orbits are estimated by assuming band 1 ($\eta$, at $\Gamma$) and 3 ($\gamma$, at $R$) are both spheres and band 2 ($\beta$, at $\Gamma$) is cubic. The error of estimated frequencies of spherical orbits at $\Gamma$ and R is within 10\% compared to the direct calculation of intersection area. However the error is about 20 \% for the cube at $\Gamma$ at $E_F$=20 meV, reflecting significant deviation from the ideal cubic geometry as the chemical potential is approached. The FS projected onto the first Brillouin zone shows the tubes along $\Gamma-X$ that are characteristic of the flat-band disappear above 15 meV (see Fig.~\ref{figs new FS}), and the size of the hole and electron pockets get smaller. This agrees with our experimental determination of FS with the $E_F$ at 30 meV.

The cyclotron effective mass is estimated using;

\begin{equation}
m^*\left(E, \hat{B}, k_{\hat{B}}\right) = \frac{\hbar^2}{2\pi} \cdot \frac{\partial}{\partial E} A\left(E, \hat{B}, k_{\hat{B}}\right)
\end{equation}
where, $A\left(E, \hat{B}, k_{\hat{B}}\right)$ is the cross-sectional area of the FS.

LL-fan diagrams (see Fig.~\ref{figs3}) were plotted to investigate the phase accumulation around the cyclotron orbit for $F_{\alpha 1}$. The peaks (n) and valleys (n+1/2) are indexed as described in the main text. From the linear fits to the data, phase information and the corresponding frequency of the extremal orbits were extracted and are presented in the main text.

In Fig. ~\ref{figs4}, a narrow frequency region around $F_{\gamma}$ is shown. Here, the arrow points to the position of the potential frequencies arising from magnetic breakdown orbits as discussion in the main text.


\begin{thebibliography}{11}
\bibitem{pd.theory} K.-H. Ahn, W. E. Pickett, and K.-W. Lee, Phys. Rev. B {\bf 98}, 035130 (2018).

\bibitem{yan2012topological} B. Yan and S.-C. Zhang, Reports on Progress in Physics {\bf 75}, 096501 (2012).

\bibitem{ando2015topological} Y. Ando and L. Fu, Annu. Rev. Condens. Matter Phys. {\bf 6}, 361 (2015).

\bibitem{PhysRevX.6.031003} Z. Zhu, G. W. Winkler, Q. Wu, J. Li, and A. A. Soluyanov, Phys. Rev. X {\bf 6}, 031003 (2016).

\bibitem{bradlyn2016beyond} B. Bradlyn, J. Cano, Z. Wang, M. Vergniory, C. Felser, R. J. Cava, and B. A. Bernevig, Science {\bf 353}, aaf5037 (2016).

\bibitem{wang2016hourglass} Z. Wang, A. Alexandradinata, R. J. Cava, and B. A. Bernevig, Nature {\bf 532}, 189 (2016).

\bibitem{muhlbauer2009skyrmion} S. Mu¨hlbauer, B. Binz, F. Jonietz, C. Pﬂeiderer, A. Rosch, A. Neubauer, R. Georgii, and P. Bo¨ni, Science {\bf 323}, 915 (2009).

\bibitem{topo.review.weyl} B. Yan and C. Felser, Ann. Rev. Cond. Matt. Phys. {\bf 8}, 337 (2017).

\bibitem{topo.review.TSC} M. Leijnse and K. Flensberg, Semicond. Sci. and Tech. {\bf 27}, 124003 (2012).

\bibitem{Pd.first} N. J. Ghimire, M. A. Khan, A. S. Botana, J. S. Jiang, and J. F. Mitchell, Phys. Rev. Materials {\bf 2}, 081201 (2018).

\bibitem{Pd2} K. Wei, K.-W. Chen, J. N. Neu, Y. Lai, G. L. Chappell, G. S. Nolas, D. E. Graf, Y. Xin, L. Balicas, R. E. Baumbach, and T. Siegrist, Phys. Rev. Materials {\bf 3}, 041201 (2019).

\bibitem{singleton2004national} J. Singleton, C. Mielke, A. Migliori, G. Boebinger, and A. Lacerda, Physica B: Condensed Matter {\bf 346}, 614 (2004).

\bibitem{supp} See Supplemental Material for (experimental detail, electronic structure calculation, electronic structure calculation, band properties with respect to chemical potential and LL-fan diagrams which includes Ref[11,14-19]). 

\bibitem{PDO.1} S. Ghannadzadeh, M. Coak, I. Franke, P. Goddard, J. Singleton, and J. L. Manson, Rev. Sci. Ins. {\bf 82}, 113902 (2011).

\bibitem{PDO.2} M. Altarawneh, C. Mielke, and J. Brooks, Rev. Sci. Ins. {\bf 80}, 066104 (2009).

\bibitem{PDO.3} P. C. Ho, J. Singleton, P. A. Goddard, F. F. Balakirev, S. Chikara, T. Yanagisawa, M. B. Maple, D. B. Shrekenhamer, X. Lee, and A. T. Thomas, Phys. Rev. B {\bf 94}, 205140 (2016).

\bibitem{ho2007haas} P.-C. Ho, J. Singleton, M. Maple, H. Harima, P. Goddard, Z. Henkie, and A. Pietraszko, New J. Phys. {\bf 9}, 269 (2007).

\bibitem{vasp} G. Kresse and D. Joubert, Phys. Rev. B {\bf 59}, 1758 (1999).

\bibitem{gga} J. P. Perdew, K. Burke, and M. Ernzerhof, Phys. Rev. Lett. {\bf 77}, 3865 (1996).

\bibitem{burkov2016topological} A. Burkov, Nature Materials {\bf 15}, 1145 (2016).

\bibitem{Berry.phase.TaP} J. Hu, J. Liu, D. Graf, S. Radmanesh, D. Adams, A. Chuang, Y. Wang, I. Chiorescu, J. Wei, L. Spinu, et al., Scientiﬁc reports {\bf 6}, 18674 (2016).

\bibitem{Berry.phase.3} D. Shoenberg, Magnetic Oscillations in Metals (Cambridge University Press, 2009)

\bibitem{BiPd.me} M. A. Khan, D. E. Graf, I. Vekhter, D. A. Browne, J. F. DiTusa, W. A. Phelan, and D. P. Young, Phys. Rev. B {\bf 99}, 020507(R) (2019).

\bibitem{AuBe} D. J. Rebar, S. M. Birnbaum, J. Singleton, M. Khan, J. C. Ball, P. W. Adams, J. Y. Chan, D. P. Young, D. A. Browne, and J. F. DiTusa, Phys. Rev. B {\bf 99}, 094517 (2019).

\bibitem{LL.fan} Y. Ando, J. Phys. Soc. Jpn. {\bf 82}, 102001 (2013).

\bibitem{Berry.phase.trivial} M. N. Ali, L. M. Schoop, C. Garg, J. M. Lippmann, E. Lara, B. Lotsch, and S. S. Parkin, Science Advances {\bf 2}, e1601742 (2016).

\bibitem{PhysRevLett.115.226401}Z. J. Xiang, D. Zhao, Z. Jin, C. Shang, L. K. Ma, G. J. Ye, B. Lei, T. Wu, Z. C. Xia, and X. H. Chen, Phys. Rev. Lett. {\bf 115}, 226401 (2015).

\bibitem{wang2018vanishing} J. Wang, J. Niu, B. Yan, X. Li, R. Bi, Y. Yao, D. Yu, and X. Wu, Proceedings of the National Academy of Sciences of the U.S.A. {\bf 115}, 9145 (2018).

\bibitem{Berry.phase.trivial2} Y. Liu, X. Yuan, C. Zhang, Z. Jin, A. Narayan, C. Luo, Z. Chen, L. Yang, J. Zou, X. Wu, et al., Nature Comm. {\bf 7}, 12516 (2016).

\bibitem{Berry.phase.2} H. Murakawa, M. Bahramy, M. Tokunaga, Y. Kohama, C. Bell, Y. Kaneko, N. Nagaosa, H. Hwang, and Y. Tokura, Science {\bf 342}, 1490 (2013).

\bibitem{PhysRevLett.113.246402} L. P. He, X. C. Hong, J. K. Dong, J. Pan, Z. Zhang, J. Zhang, and S. Y. Li, Phys. Rev. Lett. {\bf 113}, 246402 (2014).

\bibitem{Berry.phase.trivial3} J. Cao, S. Liang, C. Zhang, Y. Liu, J. Huang, Z. Jin, Z.-G. Chen, Z. Wang, Q. Wang, J. Zhao, et al., Nature Comm. {\bf 6}, 7779 (2015).

\end{thebibliography}

\begin{thebibliography}{11}
\bibitem{PDO.1s} S. Ghannadzadeh, M. Coak, I. Franke, P. Goddard, J. Singleton, and J. L. Manson, Rev. Sci. Ins. {\bf 82}, 113902 (2011).

\bibitem{PDO.3s} P. C. Ho, J. Singleton, P. A. Goddard, F. F. Balakirev, S. Chikara, T. Yanagisawa, M. B. Maple, D. B. Shrekenhamer, X. Lee, and A. T. Thomas, Phys. Rev. B {\bf 94}, 205140 (2016).

\bibitem{PDO.2s} M. Altarawneh, C. Mielke, and J. Brooks, Rev. Sci. Ins. {\bf 80}, 066104 (2009).

\bibitem{ho2007haass} P.-C. Ho, J. Singleton, M. Maple, H. Harima, P. Goddard, Z. Henkie, and A. Pietraszko, New J. Phys. {\bf 9}, 269 (2007).
\bibitem{Pd2s} K. Wei, K.-W. Chen, J. N. Neu, Y. Lai, G. L. Chappell, G. S. Nolas, D. E. Graf, Y. Xin, L. Balicas, R. E. Baumbach, and T. Siegrist, Phys. Rev. Materials {\bf 3}, 041201 (2019).
\bibitem{vasps} G. Kresse and D. Joubert, Phys. Rev. B {\bf 59}, 1758 (1999).

\bibitem{ggas} J. P. Perdew, K. Burke, and M. Ernzerhof, Phys. Rev. Lett. {\bf 77}, 3865 (1996).
\end{thebibliography}
\end{document}